\documentclass[sigconf]{acmart}
\AtBeginDocument{%
  }
\usepackage{array}
\usepackage{multirow} 
\usepackage{colortbl} 
\usepackage{float}

\setcopyright{acmlicensed}
\copyrightyear{2025}
\acmYear{2025}
\acmConference[SIGIR '25]{Make sure to enter the correct
 conference title from your rights confirmation email}{July 13--17,
 2025}{Padova, Italy}

\begin{document}


\title{PreQRAG - Classify and Rewrite for Enhanced RAG}

\author{Damian Martinez}
\affiliation{%
  \institution{University of Delaware}
  \city{Newark}
  \state{Delaware}
  \country{USA}
}
\email{damianm@udel.edu}

\author{Catalina Riano}
\affiliation{%
  \institution{University of Delaware}
  \city{Newark}
  \state{Delaware}
  \country{USA}
}
\email{criano@udel.edu}

\author{Hui Fang}
\affiliation{%
  \institution{University of Delaware}
  \city{Newark}
  \state{Delaware}
  \country{USA}
}
\email{hfang@udel.edu}

\renewcommand{\shortauthors}{Martinez et al.}

\begin{abstract}
This paper presents the submission of the UDInfo team to the SIGIR 2025 LiveRAG Challenge. We introduce PreQRAG, a Retrieval Augmented Generation (RAG) architecture designed to improve retrieval and generation quality through targeted question preprocessing. PreQRAG incorporates a pipeline that first classifies each input question as either single-document or multi-document type. For single-document questions, we employ question rewriting techniques to improve retrieval precision and generation relevance. For multi-document questions, we decompose complex queries into focused sub-questions that can be processed more effectively by downstream components. This classification and rewriting strategy improves the RAG performance. Experimental evaluation of the LiveRAG Challenge dataset demonstrates the effectiveness of our question-type-aware architecture, with PreQRAG achieving the preliminary second place in Session 2 of the LiveRAG challenge.
\end{abstract}

\maketitle

\section{Introduction}

Retrieval-Augmented Generation (RAG) has emerged as a powerful technique to enhance the performance of large language models (LLMs) by incorporating information from external knowledge sources. When grounding responses in retrieved documents, RAG systems can provide more accurate, up-to-date, and contextually relevant answers while significantly reducing hallucinations, that is, factually incorrect outputs produced by LLMs in the absence of sufficient knowledge. 

The SIGIR 2025 LiveRAG Challenge provides an open platform for academic and industry teams to explore and evaluate innovations in RAG-based question answering. The challenge is structured around a fixed corpus Fineweb \cite{penedo2024the}, and using the open-source language model Falcon-3B-10B-Instruct \cite{falcon3}. Participants were given a set of 500 questions and a strict two-hour window to generate responses, including the corresponding supporting documents from the corpus.

Our proposed solution introduces a structured pipeline that begins with query preprocessing step, where each question is classified based on whether it requires information from a single document or multiple documents. Depending on this classification, the question is then rewritten to optimize retrieval effectiveness. This reformulation step is designed to improve performance for both sparse and dense retrieval models, which are run to collect relevant documents. The retrieved results from both methods are then merged and passed through a reranking stage to prioritize the most relevant documents. Finally, a fixed number of top-ranked documents are integrated into a prompt, which is used to LLM in generating a coherent and accurate answer.


\section{System Design and Implementation}
\label{sec:system}
\begin{figure}
    \centering
    \includegraphics[width=1\linewidth]{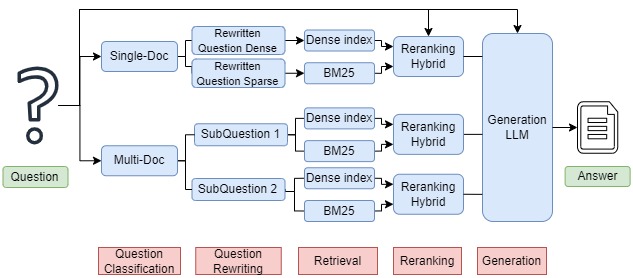}
    \caption{PreQRAG System Architecture}
    \label{fig:PreQRAG AR}
\end{figure}

PreQRAG is a comprehensive RAG system architecture designed to support LLMs in answering both single-document and multi-document questions \cite{lewis2021retrievalaugmentedgenerationknowledgeintensivenlp}. Our approach is based on the RAG paradigm \cite{gao2024retrievalaugmentedgenerationlargelanguage} by improving the quality of the retrieved supporting documents and improving the accuracy of the generated answers. The architecture, illustrated in Figure \ref{fig:PreQRAG AR}, consists of five main components: Question Classification, Question Rewriting, Retrieval, Reranking, and Generation. 

\subsection{Question Classification}
An important distinction in each question's nature is to address how many and which specific documents are needed to answer the question; this will determine the behavior in all the stages of the system. The question classification is the first component of PreQRAG, which, as a preprocessing step, classifies each input question into one of two categories:
\begin{itemize}
    \item \textbf{Single-document questions:} Expected to be answerable using a single source passage.
    \item \textbf{Multi-document questions:} Expected to be answerable using more than one source passage.
\end{itemize}

This classification informs downstream decisions such as how much context to retrieve, how to do the retrieval, and how to structure the prompts. Initially, we used Falcon3-10B-Instruct for this task, but it was later replaced with a lightweight rule-based classifier that demonstrated superior performance, particularly in identifying multi-document questions. 

Misclassifying a single-document question as multi-document is less problematic than the reverse. In the first case, providing extra context has little negative impact on the answer. In contrast, misclassifying a multi-document question as a single-document can result in insufficient context being retrieved, which increases the risk of producing incorrect or incomplete answers.

A combined method was also implemented, using Falcon and the rule-based approach together, but the rule-based classifier by itself consistently achieved the highest classification accuracy for multi-document questions. Table \ref{tab:class_methods} shows the accuracy for each classification method. The accuracy was calculated as the proportion of correctly classified questions (matching the true single or true multi-document label) over the total number of questions evaluated for each classification method.

\begin{table}[H]
    \centering
    \begin{tabular}{p{1.5cm}|c|c|c}
        \toprule
         \textbf{Question Type}&   \textbf{Rule Based}&  \textbf{Falcon3-10B-Ins}t& \textbf{Combined}\\
         \midrule
         Single-doc&  \textbf{0.731}&  0.822& 0.821\\
         Multi-doc&  \textbf{0.953}&  0.730& 0.711\\
         \bottomrule
    \end{tabular}
    \caption{Accuracy of Question Classification.}
    \label{tab:class_methods}
\end{table}
\vspace{-0.3cm}
\subsection{Question Rewriting}

In the question rewriting stage, our goal is to transform the input questions to improve retrieval accuracy and recall, particularly when used with dense and sparse retrieval models. This process includes correcting spelling errors and refining the phrasing of the questions to make them more compatible with retrieval mechanisms. We have implemented three distinct prompting strategies based on the query classification described in the previous section.

\subsubsection{\textbf{Single-document questions}}

For single-document questions, we prompt Falcon-3B-10B-Instruct to generate two rewritten versions of the original query, each tailored to a different retrieval strategy. The first rewriting prompt is designed to produce a version optimized for web-style search, aiming to enhance performance with BM25-based sparse retrieval. The second rewriting prompt focuses on dense retrieval, instructing the LLM to remove unnecessary premises and incorporate key terms directly into the question. This results in a more concise and focused query, improving the effectiveness of dense index retrieval models.

This is an example of the original query and its rewritten versions using our two distinct strategies:

\textbf{Original Query:} \\
\textit{``I'm developing new methods for protein analysis and wondering what specific technological improvements are being made to HDX methodology for complex protein systems?''}
    
\textbf{Sparse-Optimized Rewrite:} \\
\textit{``Recent technological advancements in HDX methodology for complex protein systems''}
    
\textbf{Dense-Optimized Rewrite:} \\
\textit{``What specific technological improvements are being made to HDX methodology for complex protein systems?''}

These rewrites aim to align the query structure with the strengths of each retrieval model, preserving natural language and search engine-friendly phrasing for sparse retrieval and maximizing semantic focus and keyword density for dense retrieval.

As shown in the evaluation results from the single-document setting (Appendix \ref{appendix:QR_single}), query rewriting significantly improved retrieval performance across both retrieval paradigms. Specifically, the sparse-optimized rewrite achieved a 13.34\% improvement in average Mean Reciprocal Rank (MRR) compared to the original queries when using the BM25-based sparse index. Similarly, the dense-optimized rewrite resulted in a 14.2\% increase in MRR over the original queries when evaluated with embedding-based dense retrieval.

    
    

\subsubsection{\textbf{Multi-document question}}
When the classification system identifies a multi-document question, we prompt Falcon3-10B-Instruct to decompose the original query into two distinct sub-questions. This decomposition enables the system to retrieve documents supporting each sub-question independently. The retrieved documents from both sub-questions are then aggregated to provide comprehensive evidence for answering the original complex query.

This is an example of an original multi-question query and its decomposed sub-questions:

\textbf{Original Query:} \\
\textit{``How does Azure Firewall enable network segmentation, and what are the security compliance aspects of integrating it with Azure Key Vault Managed HSM?''}
    
\textbf{Sub-question 1:} \\
\textit{How does Azure Firewall enable network segmentation?}
    
\textbf{Sub-question 2:} \\
\textit{What are the security compliance aspects of integrating Azure Key Vault Managed HSM with Azure Firewall?}

The results in Table~\ref{tab:multidoc_decomposition} shows the effectiveness of query decomposition for multi-document questions. The percentage of ground-truth documents retrieved increased substantially at all Top-K levels. For example, using the BM25-based sparse index, Top-10 retrieval improved from 41\% to 55\%. Similarly, dense retrieval performance at Top-10 rose from 36\% to 56\%.


\subsection{Retrieval Stage}

The goal of this stage is to retrieve the most relevant documents for a given query. We used pre-built indices provided by the \textit{LiveRAG} organizers: a BM25-based sparse index and a dense index.

For the dense index, documents were segmented into sentence-based chunks of up to 512 tokens using the LlamaIndex sentence splitter \cite{Liu_LlamaIndex_2022}. Each chunk was embedded into a 512-dimensional vector using the \texttt{E5-base} embedder \cite{wang2022text}. For the sparse index, the same chunks were indexed using a BM25-based method implemented on the OpenSearch platform.

To assess retrieval effectiveness, we considered an ideal evaluation setting: retrieving the most relevant document for single-document questions, and the most relevant set of documents for multi-document questions. Since the test dataset includes the original document(s) used to generate each question, we treat these as the ground truth for relevance. While this assumption holds in most cases, it is important to note that for more general questions, valid answers may exist in additional documents as well.

We evaluated performance by checking whether PreQRAG could retrieve the original document(s) within different cutoff ranks (TOP-K). Ideally, the original document should appear at the top of the ranked list in either the dense or sparse retrieval output to provide high-quality input for the generation stage.

Table~\ref{tab:performance} presents the percentage of original documents successfully retrieved using the preprocessed queries across various TOP-K values (1, 2, 3, 10) for both dense and sparse retrieval models. We limited the evaluation to TOP-10 results, as extending beyond this range may increase recall but also introduces noise, which can degrade answer quality in the generation stage.

\begin{table}[h]
    \centering
    \begin{tabular}{|c|c|c|c|c|}
    \hline
    \multicolumn{5}{|c|}{Single-doc Questions}\\ \hline
    & Top1 & Top2 & Top3 & Top 10 \\ \hline
    Sparse (Rewritten) & 37.5\%& 47.7\%& 52.1\%& 67.4\%\\ \hline
    Dense (Rewritten)& 31.2\%& 36.2\%& 42.0\%& 52.1\%\\ \hline
 \multicolumn{5}{|c|}{Multi-doc Questions}\\\hline
 & Top1 & Top2 & Top3 &Top 10 \\\hline
 Sparse (Rewritten)
& 32\%& 36\%& 40\%&55\% \\\hline
 Dense (Rewritten)
& 26\%& 30\%& 36\%&56\%\\\hline
    \end{tabular}
  
    \caption{Retrieval Performance Metrics for Single-doc and Multi-doc Questions}
    \label{tab:performance}
\end{table}
\vspace{-0.8cm}
During this analysis, we observed that some questions were better served by either the dense or the sparse retrieval model, depending on their structure and content. To leverage the strengths of both approaches, we adopted a hybrid retrieval strategy, combining the results from both dense and sparse models. The merged set of candidate documents was then passed to the reranking stage for reordering and further relevance assessment.

\subsection{Reranking Stage}

The reranking stage is responsible for refining the set of candidate documents retrieved during the initial retrieval phase. While the sparse and dense retrievers provide a broad set of potentially relevant documents, reranking enables the system to reassess and reorder these candidates using a more precise scoring function. The objective is to improve the overall ranking by leveraging a model trained specifically for fine-grained relevance estimation.

For this step, we employed \texttt{bge-reranker-v2} \cite{chen2024bge}, a cross-encoder model that jointly encodes both the question and each candidate document to produce highly accurate relevance scores. Unlike the initial retrieval models, which operate independently over question and document embeddings, the cross-encoder architecture evaluates the semantic match between the full question and document pair, enabling more nuanced relevance judgments.

We applied the reranker to the combined candidate sets retrieved from both the dense and sparse indices, as introduced in the previous stage. This hybrid retrieval set was then reranked to produce the final ordered list of documents used in the answer generation stage.

To evaluate reranking performance, we used the two datasets described in Appendix~\ref{appendix:datasets}. In the single-document subset, each query was associated with one target document, while in the multi-document subset, each query had two relevant documents. We measured performance based on the percentage of original ground-truth documents successfully retrieved after reranking. The results are presented in Table~\ref{tab:performanceRera}, and can be directly compared to the initial retrieval results shown in Table~\ref{tab:performance}, highlighting the improvements achieved through reranking.

\begin{table}
    \centering
    \begin{tabular}{|c|c|c|c|c|}
    \hline
    \multicolumn{5}{|c|}{Single-doc Questions}\\ \hline
    & Top1 & Top2 & Top3 & Top 10 \\ \hline
    Hybrid + Rerank &   64.2\%&   68.7\%&   74.4\%&   85.2\%\\ \hline
 \multicolumn{5}{|c|}{Multi-doc Questions}\\\hline
 & Top1 & Top2 & Top3 &Top 10 \\\hline
 Hybrid + Rerank & 42\%& 47\%& 53\%&70\%\\\hline
    \end{tabular}
  
    \caption{Reranking Performance Metrics for Single-doc and Multi-doc Questions}
    \label{tab:performanceRera}
\end{table}

\subsection{Generation Stage}
The final stage of the pipeline involved generating answers using Falcon3-10B-Instruct, an instruction-tuned model known for its open access and strong performance in task-oriented generation tasks \cite{falcon3}. Falcon’s training objective emphasizes instruction following, making it particularly suitable for applications involving prompt-based QA generation. Moreover, its use was a requirement of the challenge.

\textbf{Generation Process:} The generation stage in PreQRAG follows the standard RAG method; a prompt containing the input question and its associated context is passed to the model, which returns an answer. This process performs well for single-document questions. However, for multi-document questions, we explored three different variations in how generation is handled to determine whether alternative strategies could yield better results. The results did not show a significant advantage over the standard method, suggesting that prompt and context design, not the generation process itself, has the greatest impact. Refer to Appendix \ref{appendix:gen-strategies} for the details and performance metrics of the three approaches explored.

\textbf{Model Parameters:} To explore the impact of decoding strategies on answer quality, we tested different generation parameters. We found that deterministic decoding consistently yielded higher-quality answers when context passages were reliably relevant. As a result, we adopted the default non-sampling configuration of Falcon using greedy decoding. This approach ensures stable, reproducible outputs and minimizes generation variance across different runs, also enhancing the answer being grounded in the context passage given to the model. For further details and results of the different tested parameters, refer to Appendix \ref{appendix:gen-parameters}.

\textbf{Context Length}: Another important part to consider is the effect of context length on generation outcomes, by context length we mean the number of documents provided as input to the model. While it might seem intuitive that more context leads to better answers, this is not always the case. 

The results, presented in Table \ref{tab:context_gen}, show that adding more documents can sometimes degrade performance by introducing irrelevant, redundant, or conflicting information, which can reduce answer accuracy or conciseness. 

For single-document questions, using only the TOP-1 retrieved document yielded the best overall performance, which is expected if the retrieved document is correct. However, because the top document is not always the expected one, we opted for a TOP-3 context setting in the PreQRAG final configuration. This provides a better chance of including the relevant document while maintaining similar performance to TOP-1. For multi-document questions, TOP-1 performed the worst across all metrics, while TOP-3, TOP-5, and TOP-10 yielded comparable results. Given the trade-off between performance and computational efficiency, we selected TOP-3 as the optimal context length for multi-document generation as well. Please note that by "best overall performance," we mean that each metric is considered equally important in the evaluation.

\begin{table}[h]
    \centering
    \begin{tabular}{|c|c|c|c|c|}
        \hline
        Question & Context & Equivalence & Relevance & Faithfulness \\
        \hline
        \multirow{4}{*}{Single-doc} & Top1 & 0.700  &  0.927 & 0.927  \\
        \cline{2-5}
         & Top3 & 0.731 & 0.931 & 0.854  \\
        \cline{2-5}
         & Top5 & 0.722 & 0.936  & 0.863  \\
        \cline{2-5}
         & Top10 &  0.718 & 0.936  & 0.890 \\
        \hline
        \multirow{4}{*}{Multi-doc} & Top1 & 0.523 & 0.830 & 0.726 \\
        \cline{2-5}
         & Top3 & 0.604 & 0.915 & 0.820  \\
        \cline{2-5}
         & Top5 & 0.617 & 0.901  & 0.849  \\
        \cline{2-5}
         & Top10 & 0.613  & 0.905  & 0.820  \\
        \hline
    \end{tabular}
    \caption{Context Length Performance Metrics.}
    \label{tab:context_gen}
\end{table}

\vspace{-0.8cm}

\section{Evaluation}
\label{sec:evaluation}
\subsection{Metrics}
To assess the effectiveness of PreQRAG and guide design decisions, we conducted a comprehensive evaluation process that included an extra metric beyond the baseline framework proposed by the LiveRAG Challenge organizers. While the official evaluation focused primarily on correctness and faithfulness, we divided the correctness metric to address two different dimensions: equivalence and relevance. This broader set of evaluation metrics provided a clearer understanding of the model behavior and allowed us to better diagnose the effects of different configurations and methods.

\textbf{Equivalence:} Captures semantic similarity to expected answers, by comparing the generated answer to the gold answer. It uses a 3-point evaluation scale (fully, partially, or not equivalent).
    
\textbf{Relevance:} Measures if the answer addressed the question asked, independent of correctness or grounding. It uses a 3-point evaluation scale (fully, partially, or not relevant).
    
\textbf{Faithfulness:} Assess whether the generated answers were strictly grounded in the retrieved context. It uses a 3-point scale (fully, partially, or not supported). This remained the same as the metric provided by the organizers.

The evaluation was done automatically with model assistance using GPT-4o Mini, prompting it to evaluate each metric per question and give a numeric value. The individual scores were then summed across all questions for each metric and divided by the maximum possible score to obtain the final normalized metric value. GPT-4o Mini model was used for all performance metrics reported in this paper due to its significantly lower cost while maintaining reliable evaluation quality. Table \ref{tab:context_gen} shows the final performance of the system for different context scenarios, as explained in the previous section. For detailed descriptions of the evaluation prompts, see Appendix \ref{appendix:eval1}, and for comparison between the evaluation done by Claude 3.5 Sonnet and GPT-4o Mini refer to Appendix \ref{appendix:eval2}.

\subsection{Datasets}
Multiple synthetic datasets tailored to diverse question-answer scenarios were generated to evaluate the performance and generalizability of our system. These datasets were created using DataMorgana \cite{filice2025generatingdiverseqabenchmarks} a powerful tool designed for generating controlled, large-scale, and diverse QA benchmarks. DataMorgana provides an interface to define categorizations of question types, formulate rich distributions over categories within those categorizations, and generate high-quality question-answer pairs grounded in the FineWeb corpus \cite{penedo2024the}.

Following best practices and examples outlined by the authors, we adopted the categorization schema proposed in their work. Each dataset variation differs in the probability distribution applied to these categories, allowing us to control for the complexity, naturalness, and reasoning type involved. This strategy ensures our system is evaluated against a representative spectrum of question types and user profiles. Refer to Appendix \ref{appendix:datasets} for details on the datasets generated.

\section{Results}
\label{sec:results}
Table \ref{tab:liverag_results} shows the official preliminary results on the day of the LiveRAG Challenge. PreQRAG achieved a preliminary second place in Session 2, demonstrating the effectiveness of its question-type-aware architecture. In addition to the challenge ranking, we evaluated the system performance using our evaluation framework. The evaluation showed an equivalence score of 0.977 and a relevance score of 0.884. The faithfulness metric could not be computed due to the absence of gold answers. However, the results, both from the official ranking and from the internal evaluation, highlight the potential of the PreQRAG question-oriented approach as a promising strategy to improve retrieval-augmented generation systems.

\begin{table}[H]
    \centering
    \begin{tabular}{|c|l|c|c|}
        \hline
        Rank & Team Name & Correctness [-1:2] & Faithfulness [-1:1] \\
        \hline
        1 & Magikarp & 1.231578 & 0.656464 \\ \hline
        \rowcolor{lightgray}
        2 & UDInfo & 1.200586 & 0.623175 \\ \hline
        3 & RAGtifier & 1.134454 & 0.552365 \\\hline
        4 & HLTCOE & 1.070111 & 0.340711 \\ \hline
        \hline
    \end{tabular}
    \caption{LiveRAG Challenge Session 2 Results. UDInfo performance highligted in grey.}
    \label{tab:liverag_results}
\end{table}
\vspace{-0.9cm}
\section{Conclusion}
\label{sec:conclusions}

PreQRAG’s question-type awareness demonstrates promising results, showing that classifying and rewriting questions significantly enhances the retrieval component of a RAG system. During the retrieval stage, we observed that certain questions are better represented using either dense or sparse retrieval methods; therefore, a hybrid approach proved to be more effective, yielding higher retrieval efficiency. In the generation stage, the design of prompts and context arrangement plays a critical role in answer quality. This insight is particularly valuable when using a less capable model in terms of the number of parameters, as thoughtful prompt engineering and careful selection of context length can enhance performance.

\bibliographystyle{ACM-Reference-Format}
\bibliography{reference}

\section*{Appendix}
\label{sec:appendix}
\addcontentsline{toc}{section}{Appendix}
\appendix
\renewcommand{\thetable}{A.\arabic{table}}
\setcounter{table}{0}

\section{Questions Rewriting Results}
\subsection{Single-document questions}
\label{appendix:QR_single}
To assess the effectiveness of the query rewriting strategies for single-document questions, we used Mean Reciprocal Rank (MRR) as the primary evaluation metric. MRR quantifies retrieval quality by computing the inverse of the rank at which the first relevant document is retrieved, averaged over all queries. The evaluation was performed using the single-document dataset refer to Appendix \ref{appendix:datasets}. Results are presented in Table~\ref{tab:eval_rewq}, comparing baseline queries against their rewritten versions under both sparse and dense retrieval settings.

\begin{table}[h]
    \centering
    \begin{tabular}{|l|c|c|c|c|c|}
        \hline
        Query Type            & \multicolumn{4}{c|}{Top-K}             & Average \\ \cline{2-5}
                              & @1   & @2   & @5   & @10           &         \\ \hline
        Sparse (Original)     & 0.025 & 0.077 & 0.171 & 0.370          & 0.161   \\ \hline
        Sparse (Rewritten)    & 0.033 & 0.083 & 0.201 & 0.412          & 0.182   \\ \hline
        Dense (Original)      & 0.023 & 0.068 & 0.141 & 0.326          & 0.139   \\ \hline
        Dense (Rewritten)     & 0.025 & 0.073 & 0.163 & 0.374          & 0.159   \\ \hline
    \end{tabular}
    \caption{Mean Reciprocal Rank (MRR) for original and rewritten queries in the single-document setting}
    \label{tab:eval_rewq}
\end{table}

\subsection{Multi-document questions}
\label{appendix:QR_multi}

To evaluate the effectiveness of query decomposition for multi-document questions, we measured the percentage of ground-truth documents retrieved at various Top-K thresholds. This evaluation compares the baseline approach treating the  multi-document question as a single-doc question against our decomposition strategy, where each question is split into two sub-questions. The evaluation was conducted using the multi-document subset of the dataset, as detailed in Appendix~\ref{appendix:datasets}. Results are shown in Table~\ref{tab:multidoc_decomposition}, highlighting the impact of decomposition across both sparse and dense retrieval models.

\begin{table}[h]
    \centering   
    \begin{tabular}{|p{1.2cm}|p{1.6 cm}|c|c|c|c|c|}
    \hline
    \textbf{Retrieval Method}& \textbf{Approach}& \multicolumn{5}{|c|}{Top-k} \\ \cline{3-7}
        
         &  & @1& @2& @3& @5& @10\\ \hline
        Sparse & Single-doc & 16\%           & 21\%           & 26\%   & 30\%         & 41\%            \\ \hline
        Sparse                    & Multi-doc (2 subquestions) & 32\%           & 36\%           & 40\%      & 48\%     & 55\%            \\ \hline
        Dense                     & Single-doc              & 12\%           & 19\%           & 25\%  & 31\%            & 36\%            \\ \hline
        Dense                     & Multi-doc (2 subquestions) & 26\%           & 30\%           & 36\%        & 46\%   & 56\%            \\ \hline
    \end{tabular}
    \caption{Percentage of ground-truth documents retrieved for multi-document questions, comparing single-doc and multi-doc approach.}
    \label{tab:multidoc_decomposition}
\end{table}

\renewcommand{\thetable}{B.\arabic{table}}
\setcounter{table}{0}

\section{Generation Stage Results}

\subsection{Generation Process}
\label{appendix:gen-strategies}
Three methods for the multi-document question generation process were tested:
\begin{itemize}
    \item \textbf{Method 1 - Standard approach:} The retrieved documents are concatenated and used as context.
    \item \textbf{Method 2 - Two-step Context:} The question is split into sub-questions, intermediate answers are generated, then used as context for a final answer.
    \item \textbf{Method 3 - Two-step Answer:} The question is split into sub-questions, intermediate answers are generated, and these are simply merged without a final generation step.
\end{itemize}
Table \ref{tab:multidoc-gen} shows the performance metric of each method under different context sizes, top 1, top 2, and top 3 of the retrieved documents; the standard method demonstrated the best performance in general, taking into account that each metric is given the same importance.

\begin{table}[h]
    \centering
    \begin{tabular}{|c|c|c|c|}
    \hline
    \multicolumn{4}{|c|}{Top1}\\ \hline Method
    & Equivalence & Relevance & Faithfulness \\ \hline
    Method 1 & 0.60 & 0.88 & 0.14  \\ \hline
    Method 2 & 0.58 & 0.92 & 0.10  \\ \hline
    Method 3 & 0.55 & 0.92 & 0.12  \\ \hline
    \end{tabular}
    \vspace{0.3cm}
    
    \begin{tabular}{|c|c|c|c|}
    \hline
    \multicolumn{4}{|c|}{Top2}\\ \hline Method
    & Equivalence & Relevance & Faithfulness \\ \hline
    Method 1 & 0.63 & 0.93 & 0.32  \\ \hline
    Method 2 & 0.58 & 0.94 & 0.26  \\ \hline
    Method 3 & 0.52 & 0.94 & 0.17  \\ \hline
    \end{tabular}
     \vspace{0.3cm}
     
    \begin{tabular}{|c|c|c|c|}
    \hline
    \multicolumn{4}{|c|}{Top3}\\ \hline Method
    & Equivalence & Relevance & Faithfulness \\ \hline
    Method 1 & 0.64 & 0.93 & 0.52  \\ \hline
    Method 2 & 0.55 & 0.93 & 0.40  \\ \hline
    Method 3 & 0.34 & 0.57 & 0.33  \\ \hline
    \end{tabular}
    \caption{Multi-Doc Generation Process Performance Metrics on Different Methods and Context Size.}
    \label{tab:multidoc-gen}
\end{table}

\subsection{Model Parameters}
\label{appendix:gen-parameters}
To explore the impact of decoding strategies on answer quality, we tested different generation parameters that are the commonly used in generation to balance between creative fluency and factual precision, including: 
\begin{itemize}
    \item Temperature: Controls randomness by scaling the logits before sampling. Lower values make outputs more deterministic, while higher values increase diversity.
    \item Top-k sampling: Restricts token selection to the top-k most probable tokens, promoting focused yet flexible output generation.
    \item Top-p sampling: Choose tokens from the smallest probability mass p that exceeds the threshold, dynamically adapting the token pool.
\end{itemize}

Table \ref{tab:parameters}, shows the performance of four different configurations of the decoding parameters, supporting our decision to use the default non-sampling configuration on Falcon, because it has the better performance overall. These results were obtained under two conditions: an ideal context scenario, where the model was provided with the expected (gold) document, and a TOP-1 retrieval scenario, where the context consisted of the first document returned by the retrieval system.

\begin{table}[h]

    \centering
    \begin{tabular}{|c|c|c|c|}
    \hline
    \multicolumn{4}{|c|}{Ideal Context}\\ \hline Method Configuration & Equivalence & Relevance & Faithfulness \\
        \hline
        Non-sampling & 0.924 & 0.982 & 0.988 \\
        \hline
        temp=0.7 & & & \\
        top\_p=0.85 & 0.919 & 0.982 & 0.994 \\
        top\_k=20 & & & \\
        \hline
        temp=0.9 & & & \\
        top\_p=0.9 & 0.919 & 0.979 & 0.988 \\
        top\_k=40 & & & \\
        \hline
        temp=1.2 & & & \\
        top\_p=0.95 & 0.910 & 0.979 & 0.971 \\
        top\_k=60 & & & \\ \hline
    \end{tabular}

    \begin{tabular}{|c|c|c|c|}
    \hline
    \multicolumn{4}{|c|}{Top1}\\ \hline Method Configuration & Equivalence & Relevance & Faithfulness \\
        \hline
        Non-sampling & 0.739 & 0.965 & 0.890 \\
        \hline
        temp=0.7 & & & \\
        top\_p=0.85 & 0.731 & 0.959 & 0.884 \\
        top\_k=20 & & & \\
        \hline
        temp=0.9 & & & \\
        top\_p=0.9 & 0.731 & 0.965 & 0.884 \\
        top\_k=40 & & & \\
        \hline
        temp=1.2 & & & \\
        top\_p=0.95 & 0.728 & 0.959 & 0.872 \\
        top\_k=60 & & & \\ \hline
    \end{tabular}
    \caption{Model Parameters Performance with Different Configurations.}
    \label{tab:parameters}
\end{table}

\subsection{Prompt Strategy:}
\label{appendix:prompt}

Prompt design plays a central role in guiding the model’s response behavior. For that reason, we evaluated four distinct prompting strategies:

\textbf{\textit{Prompt A - Direct Prompt:}} Simply asks the model to perform a task without additional context or framing.

\textbf{\textit{Prompt B - Role-Based Prompt:}} Assigns the model a specific persona to influence its response style and reasoning.

\textbf{\textit{Prompt C - Output-Constrained Prompt:}} Explicitly specifies the required structure or expected format of the output.

\textbf{\textit{Prompt D - Chain-of-Thought Prompt:}} Encourages the model to show the step by step reasoning on the given output. 

Testing across multiple question types revealed that role-based prompting, that is Prompt B, had the best performance overall compared to the other strategies in terms of relevance, clarity, and alignment with the context, as shown in Table \ref{tab:prompt}. This is particularly effective with Falcon Instruct, which benefits from clearly defined role conditioning due to its supervised instruction tuning. By defining the model's purpose, role-based prompts reduce ambiguity and help guide inference in a more controlled manner.

\begin{table}[h]
    \centering
    \begin{tabular}{|c|c|c|c|}\hline
         Prompt Type &  Equivalence &  Relevance &  Faithfulness \\\hline
         Prompt A &  0.890&  0.971&  0.994\\\hline
         Prompt B & 0.924 & 0.982  & 0.998 \\\hline
         Prompt C &  0.910 & 0.979 & 0.994 \\ \hline
         Prompt D &  0.904 & 0.976 & 0.988 \\ \hline
    \end{tabular}
    \caption{Prompts Performance Metrics on an Ideal Context Scenario.}
    \label{tab:prompt}
\end{table}

The following are the exact prompts used on the generation stage to evaluate the performance of each Prompt Type:

\textbf{A - Direct Prompt:} Answer the question using only the provided context. Do not include any information not found in the context. Be concise. Keep the answer under 200 tokens.
    
\textbf{B - Role-Based Prompt:} You are an expert assistant. Use only the retrieved context to answer the question clearly and accurately. Do not speculate. Keep the answer under 200 tokens.
    
\textbf{C - Output-Constrained Prompt:} Based only on the provided context, answer the question in 1-3 sentences. Do not include any irrelevant information. Keep the answer under 200 tokens.
    
\textbf{D - Chain-of-Thought Prompt:} Use only the context to answer the question. If reasoning is needed, do so briefly before giving a clear final answer. Keep the answer under 200 tokens. Do not use outside knowledge.

Table \ref{tab:prompt_top1} shows the performance metrics of the prompts evaluated under a TOP-1 retrieval scenario, where the context consisted of the first document returned by the retrieval system. These results reaffirm our decision, and that Prompt B has the best performance in general, taking into account that each metric is given the same importance.

\begin{table}[h]
    \centering
    \begin{tabular}{|c|c|c|c|}\hline
         Prompt Type &  Equivalence &  Relevance &  Faithfulness \\\hline
         Prompt A &  0.699&  0.939&  0.901\\\hline
         Prompt B & 0.739 & 0.965 & 0.895 \\\hline
         Prompt C &  0.710 & 0.942 & 0.919 \\ \hline
         Prompt D &  0.725 & 0.947 & 0.890 \\ \hline
    \end{tabular}
    \caption{Prompts Performance Metrics on a Top 1 Retrieved Document Scenario.}
    \label{tab:prompt_top1}
\end{table}

\renewcommand{\thetable}{C.\arabic{table}}
\setcounter{table}{0}

\section{Evaluation Results}
\subsection{Evaluation Prompts}
\label{appendix:eval1}

The following are the prompts used to perform the evaluation:
\\
\noindent \textit{\textbf{Equivalence Prompt:}}\\ You are an assistant evaluating the equivalence of a generated answer compared to the gold answer. Equivalence measures semantic similarity, not surface similarity. Use this scale: 2: Fully equivalent — expresses the same meaning. 1: Partially equivalent — some shared meaning, but incomplete or imprecise. 0: Not equivalent — meaning differs or is wrong. \\
    Question: <question> \\
    Gold Answer: < gold answer> \\
    Generated Answer:<answer> \\
    Return ONLY your evaluation like this: Equivalence: <score>
\\
\noindent \textit{\textbf{Relevance Prompt:}}\\You are an assistant evaluating the relevance of a generated answer based on the question. Relevance is defined as how directly and completely the answer addresses the question, regardless of correctness. Use this scale: 2: Fully relevant — directly answers the question. 1: Partially relevant — somewhat related, vague, or off-topic parts. 0: Not relevant — does not answer or is unrelated.\\
    Question: <question>\\
    Generated Answer: <answer>\\
    Return ONLY your evaluation like this: Relevance: <score>\\

\noindent \textit{\textbf{Faithfulness Prompt:}} \\You are an assistant evaluating the faithfulness of a generated answer based on the given context and question. Faithfulness assesses whether the response is grounded in the retrieved passages. Use this scale: 1: Full support. All answer parts are grounded. 0: Partial support. Some parts are grounded, others are not. -1: No support. All answer parts are unsupported by the context.\\
    Context: <context> \\
    Question: <question>\\
    Generated Answer: <answer>\\
    Return ONLY your evaluation like this: Faithfulness: <score>

\subsection{Evaluation Model Comparison}
\label{appendix:eval2}

Table \ref{tab:claude} shows the performance metrics done with Claude 3.5 Sonnet. When compared with the results done with GPT-4o Mini, which are reported in the paper, Claude's results have a similar trend across metrics, indicating consistent relative performance patterns. However, the overall scores were consistently lower when using Claude.

\begin{table}[h]
    \centering
    \begin{tabular}{|c|c|c|c|c|}
        \hline
        Question & Context & Equivalence & Relevance & Faithfulness \\
        \hline
        \multirow{4}{*}{Single-doc} & Top1 & 0.701  &  0.927 & 0.781  \\
        \cline{2-5}
         & Top3 & 0.718 & 0.927 & 0.818  \\
        \cline{2-5}
         & Top5 & 0.732 & 0.931  & 0.840  \\
        \cline{2-5}
         & Top10 &  0.662 & 0.925  & 0.825 \\
        \hline
        \multirow{4}{*}{Multi-doc} & Top1 & 0.528 & 0.836 & 0.126 \\
        \cline{2-5}
         & Top3 & 0.558 & 0.889 & 0.384  \\
        \cline{2-5}
         & Top5 & 0.550 & 0.830  & 0.337  \\
        \cline{2-5}
         & Top10 & 0.557 & 0.835  & 0.307  \\
        \hline
    \end{tabular}
    \caption{Context Length Performance Metrics using Claude 3.5 Sonnet.}
    \label{tab:claude}
\end{table}

\renewcommand{\thetable}{D.\arabic{table}}
\setcounter{table}{0}

\section{Datasets}
\label{appendix:datasets}
These are the categorizations implemented to generate the datasets used to evaluate the PreQRAG system, they were the same as those proposed by the challenge organizers and the authors \cite{filice2025generatingdiverseqabenchmarks}. \\

\textbf{1. Answer-Type Categorization:} This schema defines what kind of information the question seeks, and whether answering it requires access to single or multiple documents. It includes the following categories: 
\begin{itemize}
    \item \textit{Multi-aspect:} Requires reasoning over two distinct aspects of a topic, sourced from two documents.
    \item \textit{Comparison:} Involves comparing two entities or concepts using information drawn from separate documents.
    \item \textit{Factoid:} Seeks concise factual answers like names or dates.
    \item \textit{Non-factoid:} Encourages elaboration or explanation rather than short facts.
    \item \textit{Direct:} Straightforward, context-free queries.
    \item \textit{With-premise:} Includes a user-relevant or personal premise within the query.
    \item \textit{Long:} With at least seven words.
    \item \textit{First turn:} Opening questions in a conversation, self-contained.
    \item \textit{Non-first turn:} Follow-up questions, assuming prior conversational context.
\end{itemize} 

\textbf{2. Question Formulation Categorization:} This schema classifies how the question is phrased:
\begin{itemize}
    \item \textit{Natural:} Written in colloquial, conversational style.
    \item \textit{Search query:} Structured as keyword-based web queries.
\end{itemize} 

\textbf{3. User Expertise Categorization} 
This schema captures the assumed knowledge level of the user:
\begin{itemize}
    \item \textit{Expert:} Assumes familiarity with specialized terminology.
    \item \textit{Novice:} Assumes minimal domain knowledge.
\end{itemize}

Table \ref{tab:datasets} shows the datasets generated for the evaluation of the PreQRAG system.

\begin{table}[h]
    \centering
    \begin{tabular}{|c|c|c|}\hline
        Dataset & Number of QA & Parameters \\\hline
        Single-doc & 110 & Equal probabilities\\ 
         &  & for all categories. \\
         &  & Multi-document categories \\
            &  & were excluded. \\\hline
        Multi-doc & 106 & Only multi-document \\
                 &  & questions were generated, \\
         &  & with equal probability. \\\hline
    \end{tabular}
    \caption{Generated Datasets with DataMorgana.}
    \label{tab:datasets}
\end{table}

\end{document}